\documentclass{icrc}
\usepackage{times}
\usepackage{graphicx} 
\def\lesssim{\mathrel{\hbox{\rlap{\hbox{\lower4pt\hbox{$\sim$}}}\hbox{$<$}}}}
\def\gtrsim{\mathrel{\hbox{\rlap{\hbox{\lower4pt\hbox{$\sim$}}}\hbox{$>$}}}}

\newcommand{\ffrac}[2]
  {\left( \frac{#1}{#2} \right)}
\begin{document}
\title{Horizontal and Upward  Tau Airshowers in Valleys from Mountains and Space:
 Discovering UHE Neutrinos and New physics}
\author{D.Fargion}
\affil{Physics Department,INFN,Rome University 1,Italy}

\correspondence{daniele.fargion@roma1.infn.it}

\firstpage{1}
\pubyear{2001}

\maketitle
\begin{abstract}
Upward and horizontal $\tau$ Air-showers  emerging from the Earth
crust or mountain chains are the most powerful signals of Ultra
High Energy UHE neutrinos   $\nu_{\tau}$, $ \bar\nu_{\tau} $ and
$ \bar{\nu}_e $ at PeV and higher energy. The multiplicity in
$\tau$ Air-showers secondary particles,  $N_{opt} \simeq 10^{12}
(E_{\tau} / PeV)$, $ N_{\gamma} (< E_{\gamma} > \sim  10 \, MeV )
\simeq 10^8 (E_{\tau} / PeV) $ , $N_{e^- e^+} \simeq 2 \cdot 10^7
( E_{\tau}/PeV) $ , $N_{\mu} \simeq 3 \cdot 10^5 (
E_{\tau}/PeV)^{0.85}$ make easy its discover. UHE  $\nu_{\tau}$ ,
$\nu_{\tau}$  following Super Kamiokande
 evidence of neutrino flavor mixing, ($\nu_{\mu}\leftrightarrow \nu_{\tau}$),
 should be as abundant as $\nu_{\mu}$, $ \bar\nu_{\mu}$.
 Also anti-neutrino electrons, $\bar{\nu}_e$, near the Glashow W
 resonance peak, $E_{\bar{\nu_e}} = M^2_W / 2m_e \simeq 6.3 \cdot
10^{15}\, eV$ may generate $\tau$ Air-showers .
 Upward UHE $\nu_{\tau}- N $ interaction on Earth crust at horizontal edge and
from below, their consequent upward UHE $\tau$ air-showers
beaming toward high mountains, air-planes, ballons and satellites
should flash $\gamma$, $\mu$, X and Cherenkov lights toward
detectors. Such upward $\tau$ air-shower may already hit
 nearby satellite GRO gamma detectors  flashing them by short,
  hard, diluted $\gamma-$burst at the edge of BATSE   threshold.
  The $\tau$ air-shower may test the UHE neutrino interactions
  leading to additional fine-tuned test of New TeV Physics both in
  Mountain Valleys and in Upward showers.
\end{abstract}



\section{Introduction: UHE astrophysical neutrino detection
by Vertical and Horizontal detectors} Ultra High Energy $UHE$
neutrino of astrophysical origin above tens TeV might overcome
the nearby noisy signals of secondary atmospheric neutrinos.
Present and future underground cubic Kilometer detectors are
looking for the muon penetrating tracks to associate spatially to
remarkable persistent astrophysical source (AGN,SN,Microquasar)
or to rarest GRB event. Downward muons, secondary of air-showers,
are dominating and polluting the down-vertical signals; upward
muons by UHE neutrinos $\nu_{\mu}$, $ \bar\nu_{\mu}$ at low
energies (below TeV) are again polluted by atmospheric neutrinos;
higher energy neutrinos  $\nu_{\mu}$, $ \bar\nu_{\mu}$ above
$10^{13} eV $ may better probe the astrophysical neutrino, but
upward ones  unfortunately are more and more suppressed by the
Earth opacity. Upward Tau neutrinos, to be discussed later, are
less opaque, but at $10^{13} eV -10^{14} eV $, they leave shorter
tracks and are less detectable. Therefore the best strategy in
underground detectors  should consider Horizontal Underground
Arrays. For this reason we criticized present vertical tower-like
underground array detectors and we strongly suggest to consider
the construction of wide disk-like arrays finalized to Horizontal
UHE astrophysical neutrinos. At the depth h the horizontal
distances $d{(\theta)}=$:
 \begin{equation}
 |(R_{\oplus}-h) \cdot \sin\theta \pm
 \sqrt{(R_{\oplus}-h)^{2}\cdot \sin^{2}\theta + (2hR_{\oplus} - h^2 )}|
 \end{equation}

\begin{equation}
d{(\theta=0)} = \sqrt{(2 R_{\oplus}\cdot h)}\simeq
110\sqrt{\frac{h}{Km} }\cdot Km
      \end{equation}

Such distances are not too deep to suppress UHE neutrinos even at
GZK  $10^{10} eV$ energies. Therefore nearly horizontal UHE muon
traces are cleaner signature of UHE neutrino astrophysics.

Moreover  UHE   $\nu_{\tau}$ and $\bar{\nu_{\tau}}$ may be
converted and they may reach us from high energy galactic
sources, as pulsars, Supernova remnants  or galactic
micro-quasars and SGRs , as well as from powerful
 extra-galactic AGNs, QSRs or GRBs,  even at highest (GZK)
 energy because of the large galactic (Kpcs) and extreme cosmic
 (Mpcs) distances:

 \begin{equation}
 L_{\nu_{\mu} - \nu_{\tau}} = 4 \cdot 10^{-3} \,pc \left(
 \frac{E_{\nu}}{10^{16}\,eV} \right) \cdot \left( \frac{\Delta m_{ij}^2
 }{(10^{-2} \,eV)^2} \right)^{-1}
 \end{equation}

\section{ Tau Air-shower detecting  UHE neutrinos }

 These Tau air-showers are detectable in deep valleys or on front
 of large mountain chains  like Alps , Rocky Mountains,Grand Canyons,Himalaya
  and Ande; the latter chain is near
 present AUGER project (Fargion et al.) $(1999)$ and could offer an ideal lateral source of
 nearly horizontal air-shower.
 A natural valley to locate the future Array Telescope able to
 trace such EeV Tau Air-shower fluorescent  lights is the Death
 Valley in USA: its size and depth may capture EeV events.
  The mountain chains and  the air act as a fine tuning multi filter
detector: as a screen of undesirable rare but noisy horizontal
$(>70^o)$ UHECR showers (mainly electro-magnetic ones,
 Cherenkov photons, X,gamma and  muons); very rare un-explicable hadronic
 horizontal interaction by UHE secondary pion from a mountain may occur. The
 Mountain acts as a dense calorimeter for UHE$\bar{\nu_{\tau}}$
 nuclear events (three order of magnitude
 denser than air slant depth on the horizons);
  as a distance meter target correlating $\tau$ birth
 place and its  horizontal  air-shower opening origination with
 the cosmic ray energy density; as a characteristic  anti neutrino $\bar{\nu_e}$ detector
 by the extreme resonant cross section  $\bar{\nu_e} - e$ at
 Glashow peak  and the consequent fine-tuned energy (few $PeV$) shower events;
   as a very unique  source of dense muon bundles from a mountain
   by main tau hadronic air-showering.\\

 The vertical up-ward tau air-showers (by small arrival
 nadir angle) occur preferentially at low energies nearly
 transparent to the Earth ($E_{\nu} \sim 10^{15} - 10^{16} $ eV).
 The  oblique $\tau$ air showers (whose arrival directions have large
 nadir angle), may be  related also to higher energy $\nu_{\tau}$, or
 $\bar{\nu_{\tau}}$ nuclear interactions
 ($E_{\bar{\nu_{\tau}}} \geq 10^{17} - 10^{19}$ eV). Indeed these horizontal - upward UHE
 $\nu_{\tau}$ cross a smaller fraction of the Earth volume and
 consequently they suffer less absorption toward the horizon.
 Moreover the consequent ultra-relativistic  ($E_{\bar{\nu_{\tau}}} \geq 10^{17} - 10^{19}$ eV)
  tau may travel in atmosphere for few or even hundred $Kms$
   with no absorption before the decay  to the detector located at few Kms distance.
    On the contrary  the horizontal gamma, electron pairs and
     muon showers by primary (down-ward nearly horizontal)
   UHECR proton  are severely suppressed   ($\geq 10^{-3}$) after
   crossing  $(\geq 2\cdot 10^{3}\,g \cdot cm^{-2})$ slant depth, or equivalent at one
   atmosphere,  ($\geq 16 Kms $) of horizontal atmosphere target.\\

These huge horizontal or upward air-shower signals being at least
million to billion times more abundant than the original and
unique UHE $\tau$  or UHE $\mu$ track in underground Km cube
detectors are much easier to be discovered with no ambiguity.
These high energy PeVs tau air-shower are mainly of astrophysical
nature. Indeed they cannot even be produced by PeV atmospheric
neutrino secondaries born in atmospheric muon flavor and
oscillating in tau state, because their high PeV energy and their
consequent large oscillation lengths are much (hundred times)
longer  than the  Earth  diameter.

Present $\tau$ air shower is analogous to the  well-known Learned
and Pakwasa $(1995)$ "double bang" in underground neutrino
detectors. The novelty of the present "one bang in" (the rock,
the mountain, the Earth) - "one bang out" (the air) lays in the
self-triggered explosive nature of $\tau$ decay in flight and its
consequent huge amplified air shower signal  at a characteristic
few Kms distance.

\section{The UHE $\bar{\nu}_e$, $\nu_\tau$, $\bar{\nu}_\tau$  and $\tau$ interaction lenghts }

 Moreover the expected $ \nu_{\tau} $ signals, by their
secondary tau tracks at highest cosmic ray energy window $1.7\cdot
10^{21} \,eV > E_{\tau} > 1.6\cdot 10^{17} \, eV $, must exceed
the corresponding $ \nu_{\mu} $ (or muonic) ones, making UHE $
\nu_{\tau} $ above $0.1$ EeV the most probable UHE signal. Indeed,
the Lorentz-boosted tau range length grows (linearly) above muon
range, for $ E_{\tau} \geq 1.6 \cdot 10^8 GeV $; (see Fig (1)
eq.5): the tau track reaches its maxima extension, bounded not by
bremsstrahlung radiation length nor by pair production (eq. 4),
but by growing nuclear (mainly photo-nuclear) and mainly, later,
by electro-weak interactions (eq. 6), $ R_{\tau_{\max}} \simeq
191\;Km$, at energy $ E_{\tau} \simeq 3.8\cdot 10^9\;GeV$ in
water.

\begin{figure}[h]
\includegraphics[width=0.6\textwidth]{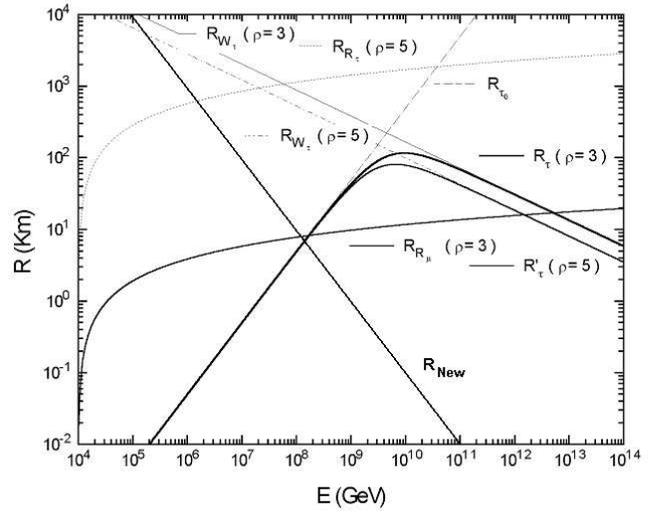} %
\caption{The tau ranges as a function of the tau energy
respectively for tau lifetime (dashed line) $R_{\tau_o}$, for
over-estimated tau radiation range $R_{R_\tau}$, (short dashed
line above) and tau electro-weak interaction range $R_{W_\tau}$,
for two densities $\rho =3$, $\rho =5$,
   $\rho_r$ (long dashed lines, continuous) and their combined range
   $R_{\tau}$. Below the corresponding radiation range
  $R_\mu$ for muons (dotted line). Finally the solid line $R_{New}$
  shows the interaction length due to New physics (extra dimension Gravity)
  at TeV for a matter density of rock $\rho =3$.}
\end{figure}

\small{
\begin{equation}\label{4}
 R_{R_{\tau}} \cong 1033 \, Km
\,\frac{5}{\rho_r}
 \left\{ 1 + \frac{\ln\left[\left(\frac{E_{\tau}}{10^8
\tiny{GeV}}\right)\left(\frac{E_{\tau}^{\min}}{10^4
\tiny{GeV}}\right)^{-1}\right]}{(\ln  10^4 )}\right\}
\end{equation}
}

\begin{equation}\label{5}
R_{\tau_o} = c \tau_{\tau} \gamma_{\tau} = 4.902 \, \mathrm{Km} \,
\left(\frac{E_{\tau}}{10^8 \, \mathrm{GeV}}\right)
\end{equation}

\begin{equation}\label{6}
 R_{W_{\tau}} = \frac{1}{\sigma N_A \rho_r} \simeq
\frac{2.6\cdot 10^3 \, \mathrm{Km}}{\rho_r} \,
\left(\frac{E_{\tau}}{10^8\, \mathrm{GeV}}\right)^{-0.363}
\end{equation}

\begin{equation} \label{7}
R_{\tiny{New}} =\frac{1}{\sigma_{\tiny{New}} N_A \rho_{r}} \simeq
 \left(\frac{E_{\tau}}{10^8
\mathrm{\tiny{GeV}}}\right)^{-1}\left(\frac{E_{\tiny{New}}^{\tiny{TeV}}}{10^3
\mathrm{\tiny{GeV}}}\right)^{-4} \frac{\tiny{Km}}{\rho_r}
\end{equation}


\begin{figure}[bt]
\begin{center}
\includegraphics[width=0.45\textwidth]
{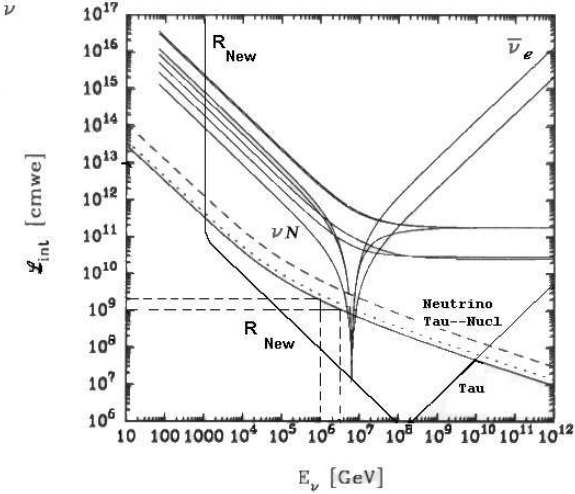}
\end{center}
  \caption {The Gandhi et all (1998) UHE neutrino
    ranges as a function of UHE neutrino energy in Earth with
    overlapping the resonant
    $\bar{\nu}_e e$, $\nu_{\tau} N$ interactions;
    below  in the corner the UHE $\tau$ range, as in Fig $1$,
     at the  same energies in matter (water).
      Finally the solid line $R_{New}$
  shows the interaction length due to New physics (extra dimension Gravity)
  at TeV for a matter density of rock $\rho =3$}
    \label{fig:boxed_graphic 5}
\end{figure}

It should be noticed that the radiative $\tau$ length estimated
above has been considered for bremsstrahlung radiation length
only. Pair production energy loss is more restrictive in the final
$R_{R_\tau}$ length (by an approximate  factor
$\frac{m_\tau}{m_\mu}$) as well as the growing photo-nuclear
interactions at highest ( tens $EeV$) energies. However the very
dominant electro-weak interactions at these energies are already
suppressing the $\tau$ growth and the combined interaction length
are slightly less, but almost comparable to the one shown in
figure above.

 At the peak maxima the tau range is nearly $10$-$20$
times longer than the corresponding muon range (at the same
energy) implying, for comparable fluxes, a ratio $10$ times larger
in $ \nu_{\tau} $ over $ \nu_{\mu} $ detection probability. This
dominance, may lead to  a few rare spectacular event a year (if
flavor mixing occurs) preferentially in horizontal plane in new
concept underground $Km^3$ detectors. The Earth opacity at those
UHE regimes at large nadir angles (nearly horizontal, few degree
upward direction) is exponentially different for UHE muons respect
to tau  above EeV : the ratio among UHE $ \nu_{\tau} $ over $
\nu_{\mu} $ tens Kms signals is exponentially high $> \exp(10)$.

 Therefore UHE Tau $ E_{\tau} \geq \ 10^5 GeV - 5 \cdot 10^7 GeV  $
air-shower in front of high mountains chains will be easily induce
peculiar horizontal UHE $\tau$ (Fargion, Aiello, Conversano 1999).
 Energies above will be probably missed.
 An hybrid detector (gamma/optical air-shower array)
 would get precise signal and arrival direction.
Because of the different neutrino interactions with energy
 it will be possible to estimate, by stereoscopic,
directional and time structure signature, the spatial air-shower
origination in air, the primary tau distance decay from the
mountain (tens  or a hundred of meters for fine tuned PeVs UHE $
\bar{\nu}_e $ and
 meters up to few Kms for  UHE $\nu_\tau$,$ \bar\nu_{\tau}$ at
 wider energy window $ E_{\tau} \geq \ 10^5 GeV - 5 \cdot 10^7 GeV $.
  Additional energy calibration may be derived sampling shower intensities.\\
   Hundreds of array (scintillator,Cherenkov) detectors
 in deep wide valley  horizontally oriented would be necessary to get tens
$\tau$ air--showers events a year; the induced $\bar{\nu}_e e
\rightarrow \tau$ air shower even in absence of $\nu_{\mu}
\leftrightarrow \nu_{\tau}$ oscillation should be well identified
and detectable. More copious ($> 5$ times more) events by PeV up
to tens PeV  charged current $\nu_{\tau} N$ interaction occur
following Super Kamiokande  flavor mixing discover.
 It will be also possible to observe UHE $\nu_{\tau}$,
by  the upward tau air-shower arriving from tens or  hundred
Kilometers away (near horizontal edges) from high mountains, high
balloon and satellites; such UHE tau created within a wide (tens
thousands to millions square km$^2$ wide and hundred meter UHE Tau
depth in Earth crust) target would discover only UHE $\nu_\tau$,$
\bar\nu_{\tau}$ neutrinos at PeV up to EeV energies and above,
just within the mysterious GZK frontiers.
 From the same highest mountains, balloons and near orbit satellite, looking more
downward toward the Earth  it is possible to discover more
frequent but lower energetic astrophysical $\simeq$ $10^{14}\div
10^{16}$ $eV$ neutrinos (by consequent Tau Air-showers) being
nearly transparent to the Earth volume;(See Fig.2).
The UHE
neutrinos $\bar{\nu_e}$,${\nu}_{\mu}$ $\bar{\nu}_{\mu}$ are
default and expected UHECR ( $\gtrsim 10^{16}$ eV) secondary
products near AGN or micro-quasars by common photo-pion decay
relics by optical photons nearby the source (PSRs, AGNs) ($p +
\gamma \rightarrow n + \pi^+, \pi^+ \rightarrow \mu^+ \nu_{\mu},
\mu^+ \rightarrow e^+ \nu_e \bar{\nu}_{\mu} $), or by proton
proton scattering in galactic interstellar matter. The maximal
observational distances from mountains, balloons or satellites,
may reach $\sim$ 110 Km $(h/Km)^{\frac{1}{2}}$ toward the horizon,
 corresponding to a UHE $\tau$ energy $\sim 2 \cdot 10^{18}$ eV $(h/ Km)^\frac{1}{2}$.
 Therefore we propose to consider such upward shower nearly horizontal detection
  to test  the highest UHE $\nu$ energies at  GZK cut off.
The expected downward muon number of events $ N_{ev} (\bar{\nu}_e
e\to\bar{\nu}_{\mu} \mu)$ in the resonant energy range, in Km$^3$,
was found to be $N_{ev} = 6$ a year. One expect a comparable
number of reactions $ (\bar{\nu}_e e\to \bar{\nu}_{\tau} \tau) $.
However the  presence of primordial $\nu_{\tau}, \,
\bar{\nu_{\tau}}$ by flavor mixing and $\nu_{\tau}, \,
\bar{\nu_{\tau}} N$ charged current interactions lead to a factor
5 larger rate, $ N_{ev} = 29$ event/year.
 If one imagines a gamma/optical detector at 5 km far in front
of a chain mountain  (nominal size 10 km, height 1 km) one finds a
$\tau$ air shower volume observable within a narrow
 beamed cone (Moliere radius $\sim 80$ m / distance $\sim 5$ Km):
 ($\Delta \theta \sim    1^o$, $\Delta \Omega \sim 2 \cdot 10^{-5}$) and
 an effective volume $V_{eff} \simeq 9 \cdot 10^{-5} $
 Km$^3$ for each observational detector.
  Each single detector is comparable to
 roughly twice a Super Kamiokande detector.
  Following  common AGN - SS91 model [The Gandhi et all,1998] 
  with a flux at present AMANDA-Baikal bounds we foresee a total event rate of:
  (6) ($\bar{{\nu}_e} e$) + (29) ($\nu_{\tau} N$) = 35 UHE $\nu_{\tau}$
  event/year/Km$^3$.
 At energies above 3 PeV we may expect a total rate of N$_{ev}$ $\sim$
 158 event/year in this mountains valley and
 nearly 3.2 $\cdot 10^{-3}$ event/year for each m$^2$ size detector.
 In a first approximation, neglecting Earth opacity,
  it is possible to show that the Earth volume observable from the top of
a mountain at height $h$, due to UHE $\tau$ at 3 PeV crossing from
below, is approximately V $\approx 5 \cdot 10^4$ Km$^3$
$\ffrac{h}{Km} \ffrac{E_{\tau}}{3\,PeV}$. These upward shower
would hit the top of the mountain. For the same $\tau$ air shower
beaming ($\Delta \theta \sim 1^o$, $\Delta \Omega \sim 2 \cdot
10^{-5}$) we derive now an effective volume $\sim$ 1 Km$^3$.
Therefore a detector open at $2 \pi$ angle on a top of a 2 Km
height mountain may observe nearly an  event every two month from
below the Earth. The gamma signal above few MeV would be
(depending on arrival nadir angle) between  $3\cdot 10^{-2}$
cm$^{-2}$ (for small nadir angle) to $10^{-5}$ cm$^{-2}$ at far
distance within 3 PeV energies. A contemporaneous (microsecond)
optical flash ($\gtrsim 300 \div 0.1 \, cm^{-2}$) must occur.
Keeping care of the Earth opacity, at large nadir angle ($\gtrsim
{60}^0$) where an average Earth density may be assumed ($< \rho >
\sim 5$) the transmission probability and creation of upward UHE
$\tau$ is approximately
\begin{equation}
P(\theta,\, E_{\nu}) = e^{\frac{-2R_{Earth} \cos
\theta}{R_{\nu_{\tau}}(E_{\nu})}} (1 - e^{-
\frac{R_{\tau}(E_{\tau})}{R_{\nu_{\tau}}(E_{\nu})}}) \, .
\end{equation}
This value, at PeV is within a fraction of a
million(${\theta}{\approx}{60}{^0}$) to a tenth of thousands
(${\theta}{\approx}{90}{^0}$). The corresponding angular integral
effective volume observable from a high mountain (or balloon) at
height $h$ (assuming a final target terrestrial density $\rho =
3$)  is:

\begin{footnotesize}
\begin{equation}
  V_{eff} \approx 0.3 \, Km^3 \ffrac{\rho}{3}\ffrac{h}{Km} e^{-
  \ffrac{E}{3\,PeV}}
  \ffrac{E}{3\,PeV}^{1.363}
\end{equation}
\end{footnotesize}
 A popular "blazar" neutrino flux model (like Berezinsky ones)
  normalized within a  flat spectra(at a standard
energy fluence$\simeq 2 \,10^{3} \frac{eV}{cm^{2}}$) is leading,
above 3 PeV, to $\sim$ 10 UHE $\nu_{\tau}$ upward event/Km$^3$
year. Therefore we must expect an average upward effective event
rate observed on a top of a mountain (h $\sim 2\,Km$) (Fig. 4):

\begin{footnotesize}
\begin{equation}
  N_{eff} \simeq 8 \, \frac{\mathrm{events}}{\mathrm{year}} \ffrac{\rho}{3} \ffrac{h}{2 \,Km}
e^{- \ffrac{E}{3\,PeV}} \ffrac{E}{3\,PeV}^{1.363}
\end{equation}
\end{footnotesize}
This rate is quite large and one expected $\tau$ air air-shower
signal (gamma burst at energies $\gtrsim 10 \, MeV$) should be
$\phi_{\gamma} \simeq 10^{-4} \div 10^{-5}$ cm$^{-2}$, while the
gamma flux at ($\sim 10^5 \, eV$) or lower energies (from electron
pair bremsstrahlung) may be two order of magnitude larger. The
optical Cherenkov flux is large $\Phi_{opt} \approx 1$ cm$^{-2}$.

\section{Upward $\tau$ Air Shower in Terrestrial Gamma Flash: evidences of UHE neutrinos?}
The tau upward air showers born in a narrow energy
 window, $10^{15}$ eV $ \lesssim E_{\nu} \lesssim  5 \cdot 10^{16} $
 eV (Fig.3) may penetrate high altitude leaving rare beamed upward gamma
 shower bursts whose sharp ($\sim $ hundreds $\mu$sec because of the hundred kms high
 altitude shower distances) time
 structure and whose hard ($\gtrsim 10^{5}$eV) spectra may hit near
 terrestrial satellites.
We claim (Fargion 2000) that such gamma upward events originated
by tau air showers produce gamma bursts at the edge of GRO-BATSE
sensitivity threshold. In particular we argue that very probably
such upward gamma events have been already detected since April
1991 as serendipitous sharp ($\lesssim 10^{-3}$ sec) and hard
($\gtrsim 10^5$ eV) BATSE gamma triggers originated from the
Earth and named consequently as Terrestrial Gamma Flashes (TGF).
The visible Earth surface from a satellite, like BATSE, at height
$h \sim 400$ Km and the consequent effective volume for UHE
$\nu_{\tau} N$ PeVs interaction and $\tau$ air shower beamed
within $\Delta \Omega \sim 2 \cdot 10^{-5} rad^2$ is: (note
$<\rho> \simeq 1.6$ because 70 \% of the Earth is covered by seas)
 $ V_{eff} = V_{TOT} \Delta \Omega \simeq 60 \, Km^3.$
The effective volume and the event rate should be reduced, at
large nadir angle ($\theta > 60^o$) by the atmosphere depth and
opacity (for a given $E_{\tau}$ energy). Therefore the observable
volume may be reduced approximately to within 15 Km$^3$ values and
the expected UHE PeV event rate is
\begin{equation}
  N_{ev} \sim 150 \cdot e^{-
  \ffrac{E_\tau}{3\,PeV}}
  \ffrac{E_\tau}{3\,PeV}^{1.363}
  \ffrac{h}{400 Km} \,
  \frac{\mathrm{events}}{\mathrm{year}}
\end{equation}
The TGF signals would be mainly $\gamma$ at flux $10^{-2}$
cm$^{-2}$ at X hundred keV energies. The observed TGF rate is
lower than the expected one (eq.~11) by nearly an order of
magnitude, and this suggests higher $E_{\nu}$ energies (to
overcome BATSE threshold) and consequently small additional
probability suppression fitting the observed TGF events rate.
 We notice that among the 75 records only 47 are published by NASA in
 their details, while 28 TGF are still unpublished. Their data
 release is therefore urgent and critical.
 While Blue Jets might be in principle triggered by upward tau
 air showers in the atmosphere (a giant "Wilson" room) we believe they are
 not themselves source of TGF. In particular their observed
 characteristic propagation velocity ($\lesssim 100$ Km/s) from
 distances $\sim$ 500 Km, disagree with short TGF millisecond
 timing and would favor a characteristic TGF time of few seconds.
  Moreover TGF data strongly dis-favor by its hard spectra
 the terrestrial Sprites connection.
 The correlations of these clustered TGFs directions
 toward  well known and maximal powerful galactic and
 extra-galactic sources either at TeV, GeV-MeV, X band , recent
 first anisotropy discovered on UHECR at EeV by AGASA, (see Hayashida 1999,Fargion
 2000), Milky Way Galactic Plane (Fig.3) and Center and well known EGRET sources,  support
 and make suggestive the TGF  identification as secondary gamma
 burst tail of UHE $\tau$ induced upward air shower.
   The present TGF-$\tau$ air-shower identification could not
    be produced by UHE $\bar{\nu}_e$ charged current
   resonant event at ($E_{\bar{\nu_{e}}} = M^2_W / 2m_e = 6.3 \cdot 10^{15}$
   eV), because of the severe Earth opacity for such resonant
   $\bar{\nu}_e$, and therefore it stand for the UHE $\nu_{\tau}
 \bar{\nu_{\tau}}$ existence. Consequently it gives support to the
 Superkamiokande evidences for $\nu_{\mu}\leftrightarrow \nu_{\tau}$
  flavor mixing from far PSRs or AGNs sources toward the Earth.
  The same argument, as shown in Fig.2, imply a new upper bound on the possible New Physics Energy edge:
  It should not  arise at threshold energies below  ($E_{New}=3 TeV$).
     At the present the very probable $\nu_{\tau}
     \bar{\nu_{\tau}}$ source of TGFs and their probable partial galactic location
     infer a first lower bound on
     $\Delta_{m_{\nu_{\mu} \nu_{\tau}}}$ ($L < 4$ Kpc, $\Delta_{m_{\nu_{\mu} \nu_{\tau}}}
      > 10^{-8}$ eV$^2$)
     and it offers a first direct test of
     the same existence of the last evanescent (hardly observed only recently),
     fundamental neutral lepton  particle: $\nu_{\tau}$ and $ \bar{\nu_{\tau}}$.
     The new physics interaction at TeV while forbid UHE signals in
     underground $Km^3$ detectors it will amplify the $\nu_{\tau}$
     signals by two order of magnitude making extremely fruit-full
     UHE $\nu_{\tau}$ astrophysics in near future.
\begin{figure}[bt]
\begin{center}
\includegraphics[width=0.45\textwidth]
{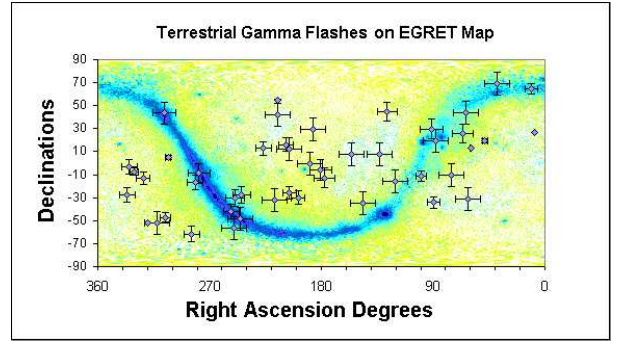}
  \caption {The TGF on EGRET map. The TGF clustering toward the galactic center and known
  EGRET sources, their squeezing along the Galactic Plane make them probably of astrophysical
  nature.}
\end{center}
    \label{fig:boxed_graphic 5}
\end{figure}
\begin{acknowledgements}
The author thanks P.G. De Sanctis Lucentini.
\end{acknowledgements}

\end{document}